\documentclass[11pt]{revtex4}
\usepackage{epsf}
\usepackage{epsfig}

\newcommand{\simgt}{\lower.5ex\hbox{$\; \buildrel > \over \sim \;$}}
\newcommand{\simlt}{\lower.5ex\hbox{$\; \buildrel < \over \sim \;$}}

\begin{document}
\title{Astrophysical Condition on the attolensing
as a possible probe for \\a modified gravity theory}

\author{Takahiro Sato${}^{\dagger}$, 
    Bobby E. Gunara${}^{\dagger\dagger}$, 
    Kazuhiro Yamamoto${}^{\dagger}$, 
    Freddy P. Zen${}^{\dagger\dagger}$
}
\affiliation{${}^{\dagger}$
Graduate School of Science, Hiroshima University, 
Higashi-Hiroshima, 735-8526, Japan\\
${}^{\dagger\dagger}$
Theoretical Physics Laboratory, Faculty of Mathematics and 
Natural Science, Institut Teknologi Bandung, Jl. 
Ganesha 10 Bandung 40132, Indonesia}
\begin{abstract}
We investigate the wave effect in the gravitational lensing 
by a black hole with very tiny mass less than 
$10^{-19} M_{\rm sun}$ (solar mass), 
which is called attolensing, 
motivated by a recent report that the lensing signature 
might be a possible probe of a modified gravity theory 
in the braneworld scenario. 
We focus on the finite source size effect and the effect 
of the relative motion of the source to the lens, which 
are influential to the wave effect in the attolensing.
Astrophysical condition that the lensed interference signature 
can be a probe of the modified gravity theory is demonstrated.
The interference signature in the microlensing system is also 
discussed.  
\end{abstract}


\maketitle

\section{Introduction}
Many physicists have drawn attention to extra dimensional physics for
several years due to recent development in testing Randall-Sundrum
type II (RS-II) scenario \cite{RSII}. In this scenario they
considered a four dimensional positive-tension brane embedded in
five dimensional AdS bulk which allows us to reconsider our
understanding about the history of our universe in the early
stages. Some investigations have been carried out to modify the
existence of primordial black holes (PBH)s in this RS-II scenario
that the life time of five dimensional PBHs against the Hawking radiation 
becomes longer compared with the standard PBH in four dimensions \cite{GCL}. 
This is because the five dimensional feature becomes significant 
in the black hole with very tiny mass.
The ratio of the life time against the Hawking radiation of such 
five dimensional PBH to that of the four dimensional PBH may be 
estimated, $lM_4/l_4M$,
where $l$ is the AdS radius of the braneworld model, $l_4$ and 
$M_4$ are the four dimensional Planck length and mass, 
respectively, and $M$ is the black hole mass \cite{GCL}.

Since the braneworld PBHs can live longer, then it is possible 
that such black holes still exist and spread out in our universe. 
Therefore, it is natural to consider the possibility of 
the gravitational lensing phenomenon by such the black hole. 
More recently, the authors \cite{KPIII} investigated the wave
effect in gravitational lensing by the black hole with the very 
tiny mass smaller than $10^{-19} M_{\rm sun}$, where 
$M_{\rm sun}$ 
is the solar mass, which is called attolensing. 
They showed that the interference signature in the energy 
spectrum in gamma ray burst due to the attolensing might 
be a possible probe of the modified gravity theory.
In the standard general relativity, PBH with the mass 
smaller $10^{-18} M_{\rm sun}$ will be evaporated through the 
Hawking radiation within the cosmic age. 
Then, the detection of such the interference signature 
would be a probe of extra dimension of our universe.

In the present paper, we consider the two effects in the lensing phenomenon
that are influential for measurement of the interference signature in the 
energy spectrum in the attolensing: One is the finite source size effect. 
The other is the effect of the relative motion of the black 
hole (lens object) to the source. We demonstrate the condition that 
these effects become influential. This is one of the astrophysical conditions 
that the attolensing can be a probe of the modified gravity theory.
Throughout this paper, $ H_0 = 72$ km s$^{-1}$ Mpc$^{-1}$ is 
the Hubble parameter, and we use the unit in which 
the light velocity equals 1.

\section{Review of Basic Equations}
The wave effect in the gravitational lensing has been 
investigated (e.g., \cite{SEF,Nakamura,path}).
We start with a brief review of the black hole solution in 
the type II Randall-Sundrum braneworld gravity model. 
We write the line element as \cite{KPIII,GT}
\begin{eqnarray}
  ds2=-(1+2\Psi)dt^2+(1+2\Phi)d{\bf x}^2,
\end{eqnarray}
where
\begin{eqnarray}
  &&\Psi=-{GM\over r}-{2\over 3}{GM l^2\over r^3},
\\
  &&\Phi=+{GM\over r}+{1\over 3}{GM l^2\over r^3},
\end{eqnarray}
where $M$ is the black hole mass, $G$ is the 
gravitational constant, and $l$ is the AdS radius of the braneworld model.
The propagation of the electromagnetic wave can 
be approximated by the massless scalar wave equation, 
which yields
\begin{eqnarray}
  \left(\triangle+\omega^2\right)\phi=(\Psi-\Phi)\phi,
\end{eqnarray}
where we assumed the amplitude of the field is in proportion to 
$e^{-i\omega t} \phi({\bf x})$, and $\omega$ is the angular 
frequency of the wave. 
With defined
\begin{eqnarray}
  && \tilde \psi=\int dr\left(\Psi-\Phi\right)
  =4GM\ln \rho-2{GMl^2\over \rho^2}+{\rm constant},
\end{eqnarray}
the amplification factor is (e.g.,\cite{SEF,Nakamura,path})
\begin{eqnarray}
F 
&=&{\omega \over 2\pi i} {D_L D_S\over D_{LS}} \int d^2\vec\theta \exp
\left[ i\omega \left({D_LD_S\over D_{LS}}
  \biggl|\vec \theta -\vec \theta_0\biggr|^2
  -\tilde \psi\right)\right],
\end{eqnarray}
where, $\vec\theta_0$ denotes the position of the source,  
$D_S$ is the (angular diameter) distance between the 
observer and the source,
 $D_L$ is the distance between the observer and the lens, and 
 $D_{LS}$ is the distance between the lens and the source.
Introducing the Einstein angle,
\begin{eqnarray}
  && \theta_E=\sqrt{4GM
 D_{LS}\over D_SD_L}, ~~ 
\end{eqnarray}
and the dimensionless variables
\begin{eqnarray}
  {\bf x}={\vec \theta \over \theta_E}, ~~ 
  {\bf y}={\vec \theta_0 \over \theta_E}, ~~ 
\end{eqnarray}
and 
\begin{eqnarray}
  && w 
  =4GM\omega, ~~  
 \varepsilon_l={l\over \theta_E D_L},
\end{eqnarray}
we have
\begin{eqnarray}
  && F={w\over 2\pi i} \int d^2 {\bf x} \exp\left[ iw\left(
  {1\over 2}|{\bf x}-{\bf y}|^2 -\ln |x| +{\varepsilon_l^2\over 2 |x|^2} 
\right) \right].
\end{eqnarray}

In the limit of the geometric optics, the light path is determined by
the lens equation,
\begin{eqnarray}
  x-y-{1\over x} - {\varepsilon_l^2\over x^3}=0.
\end{eqnarray}
{}From \cite{KPIII}, the solution is approximately obtained as
\begin{eqnarray}
  x={1\over 2} \left(\sqrt{4+y^2}\pm|y|\right)+
    {1\over 2} \left({2+y^2\over \sqrt{4+y^2}}\mp|y|\right)\varepsilon_l^2.
\end{eqnarray}
Then, the magnification is 
\begin{eqnarray}
  |\mu^\pm|={1\over 2} \left({2+y^2\over |y|\sqrt{4+y^2}}\pm 1\right)-
    {2 \over |y|\sqrt{4+y^2}^3}\varepsilon_l^2,
    \label{geom}
\end{eqnarray}
and the time delay is
\begin{eqnarray}
  {\Delta \tau}={1\over 2}|y|\sqrt{4+y^2}+\ln \left({\sqrt{ 4+y^2}+|y|\over
\sqrt{ 4+y^2}-|y|}\right)+{\varepsilon_l^2\over2} |y|\sqrt{4+y^2}.
\end{eqnarray}
In the semiclassical limit, the magnification is given by
\begin{eqnarray}
 {\cal M}(w,y)=|F|^2
\simeq{2+y^2\over |y|\sqrt{4+y^2}}+{2\over |y|\sqrt{4+y^2}}\sin(\omega\Delta\tau)-{4+2(2+y^2)\sin(w\Delta\tau)\over |y|\sqrt{4+y^2}^3}
{\varepsilon_l^2}.
\label{calmw2}
\end{eqnarray}
The first term of the right hand side of Eq.~(\ref{calmw2}) corresponds 
to the formula within the geometric optics, and the second term
does to the semiclassical correction of the wave optics, 
and the third term does to the correction due to the modification of 
the gravity.

As discussed in the literature \cite{KPIII}, one can write
\begin{eqnarray}
  \varepsilon_l^2\sim 10^{-18} \biggl({l\over 0.2 {\rm mm}}\biggr)^2
\biggl({M \over 10^{15} {\rm g}}\biggr)^{-1}
\biggl({H_0^{-1}\over D_{LS}D_L/D_S}\biggr).
\end{eqnarray}
Thus the correction is small for the primordial braneworld 
black hole. Then, in the following, 
we neglect the correction of order $\varepsilon^2_l$.

\section{Finite source size effect}
\def\bfy{{\bf y}}
In this section, we consider the finite source size effect, which 
can be influential to the interference signature \cite{MY,SPG}.
The magnification of a source with a finite size 
can be expressed by averaging  the point source 
magnification over the source-position,
\begin{eqnarray}
  \bar {{\cal M}}(w) ={\displaystyle{\int d^2 \bfy W(\bfy) {\cal M}(w,y)} 
  \over 
  \displaystyle{\int d^2 \bfy W(\bfy)}},
\label{barmw}
\end{eqnarray}
where $W(\bfy)$ denotes the distribution of the source intensity.
In this paper, 
we assume the uniform intensity of a circular region, for simplicity,
\begin{eqnarray} 
W({\bf y})= \left\{
\begin{array}{ll}
1~~~& {\rm for}~~~~|\bfy-\bfy_0|\leq R,
\\
0~~~&  {\rm for}~~~~|\bfy-\bfy_0|> R,
\end{array}
\right.
\end{eqnarray}
where $\bfy_0$ is the position of the center of 
the source and $R$ is the radius of the source.

We demonstrate the behavior of ${\bar {\cal M}}(w)$ as function of 
$w$ in several cases of $W(\bfy)$.
The panels of Figure \ref{m2_6} show $\bar {\cal M}(w)$ for the 
source distribution function $W(\bfy)$, as shown in the panels 
of Figure \ref{m2_6w}, correspondingly. 
The radius, $R$ and the center-position 
of extended source $(|\bfy_0|)$ are 
$(R,|\bfy_0|)=(0.25,0.5)$ for the panel (a), 
$(0.5,0.75)$ for the panel (b), 
$(0.75,0.25)$ for the panel (c), 
$(1.25,0.25)$ for the panel (d), respectively.
The dashed circle in the panel of Figure \ref{m2_6w} is the 
Einstein radius. 

In each panel of Figure~\ref{m2_6}, the solid curve 
is $\bar {\cal M}(w)$ defined by Eq. (\ref{barmw}), 
but the dotted straight line is the result of the geometric 
optics,
\begin{eqnarray}
  \bar {{\cal M}}_{\rm geo}={\displaystyle{\int d^2 \bfy W(\bfy) 
  \Bigl({|\mu^+(y)|+|\mu^-(y)|}}\Bigr) 
  \over 
  \displaystyle{\int d^2 \bfy W(\bfy)}}.
\label{barmwg}
\end{eqnarray}
The dashed curve in each panel of 
Figure~\ref{m2_6} shows the magnification $ {\cal M}(w,y_p)$ 
of the point source located at the position $y_p$, 
where $y_p$ is determined by solving 
\begin{eqnarray}
  \bar{\cal M}_{\rm geo}=|\mu^+(|y_p|)|+|\mu^-(|y_p|)|=
  {2+|y_p|^2\over |y_p|\sqrt{4+y_p^2}},
\label{barmpoint}
\end{eqnarray}
which yields
\begin{eqnarray}
  y_p^2=-2+\frac{2{\bar{\cal M}}_{\rm geo}}
  {\sqrt{{{\bar {\cal M}}_{\rm geo}}^2-1}}.
\label{ymu}
\end{eqnarray}
Note that the behavior of $\bar {{\cal M}}(w)$ is similar to 
$ {\cal M}(w,y_p)$ for $w\sim 1$, while 
$\bar {{\cal M}}(w)$ approaches to 
$\bar {{\cal M}}_{\rm geo}$ as $w$ becomes larger, $w\gg1$.

\section{Relative motion of lens}
In this section we consider the point source, but taking the 
relative motion of the source to the lens into account. 
Because black holes have naturally the velocity 
dispersion in the universe, then the position of the source
relative to the lens moves within a finite observation time. 
Assuming that the time resolution to obtain the energy 
spectrum is not very fine and that the energy spectrum 
is obtained by the formula (\ref{barmw}), but with 
\begin{eqnarray} 
W({\bf y})= \left\{
\begin{array}{ll}
\displaystyle{\int dt\delta^{(2)}(\bfy-\bfy(t))}
~~~& {\rm for}~~~~t_{ini}\leq t \leq t_{fin},
\\
0~~~&  {\rm for}~~~~t< t_{ini},~t> t_{fin},
\end{array}
\right.
\label{defwmotion}
\end{eqnarray}
where $\bfy=\bfy(t)$ defines the track of the source. 
Substituting (\ref{defwmotion}) into (\ref{barmw}), we have
\begin{eqnarray}
\bar{\cal M}(w)={\displaystyle{\int_{t_{ini}}^{t_{fin}}dt {\cal M}(w,y(t))}\over\displaystyle{\int_{t_{ini}}^{t_{fin}}dt}}.
\end{eqnarray}

The panels of Figure \ref{m1_7w} show the typical tracks of 
the source $W({\bf y})$ considered here. 
Figure \ref{m1_7} shows the magnification 
corresponding to the tracks.  
The track is defined by the straight line connecting
two points between $(0.5,0)$ and $(1,0)$ for the panel (a),
$(0.1,0)$ and $(0.6,0)$ for  the panel (b)
$(0.5,0)$ and $(0.5,0.75)$ for  the panel  (c)
$(0.5,0)$ and $(0.5,1.5)$ for  the panel (d), respectively.
The meaning of the solid, dotted, and dashed 
curve is same as those of Figure \ref{m2_6}.

Similar to the case of the extended source, 
$\bar {{\cal M}}(w)$ takes similar value of
$ {\cal M}(w,y_p)$, for $w\sim 1$. While $\bar {{\cal M}}(w)$
approaches to $\bar {{\cal M}}_{\rm geo}$ as $w$ becomes large, $w\gg1$.
However, note that the convergence of $\bar {{\cal M}}(w)$
  to $\bar {{\cal M}}_{\rm geo}$
in the region $w\gg1$ is slow in comparison with the case of the 
extended source in the previous section. 

\section{Condition of phase cancellation}
\def\bfy{{\bf y}}

{}From the formulas (\ref{calmw2}) and (\ref{barmw}), we expect 
that the interference signature disappears under the condition, 
\begin{eqnarray}
  w|\Delta \tau(y_{\rm min})-\Delta \tau(y_{\rm max})|\simgt 2\pi,
\label{condone}
\end{eqnarray}
where $y_{\rm min}$($y_{\rm max}$) is the minimum (maximum) value of $y$ 
in the (typical) region of the source defined by $W(\bf y)$. 
This condition means that the phase difference between the waves 
from $y_{\rm min}$ and $y_{\rm max}$ in the source plane is larger than $2\pi$.
Namely, Eq.~(\ref{condone}) is the condition of the phase cancellation. 

In the case of the point-mass lens, formula (\ref{condone})
can be approximated in a simpler formula, as follows.
Figure \ref{dtau} shows $\Delta\tau(y)$ as a function of $y$. This suggests
that $\Delta\tau=2y$ is a quite good approximation as long as 
$y\simlt1$, inside the Einstein radius. 
This is understood by the Taylor expansion of $\tau(y) $, 
\begin{eqnarray}
  \tau (y)= 2\left[y+{1\over 24}y^3-{1\over 640}y^5+{\cal O}(y^7)\right], 
\end{eqnarray}
which suggests that the correction to the relation $\tau (y)= 2y$ from 
the higher order terms is small as long as $y\simlt1$. 
Then, (\ref{condone}) is written as 
\begin{eqnarray}
  w|y_{\rm min}-y_{\rm max}|\simgt \pi.
\label{condtwo}
\end{eqnarray}

The panels of Figure \ref{m3_1} show the magnification assuming 
$W(\bfy)$ shown in each panel of Figure \ref{m3_1w}, correspondingly.
In each panel of  Figure \ref{m3_1w}, we consider a few 
case of $W(\bfy)$ with the same $|y_{\rm max}-y_{\rm min}|$ but with 
different position of the center. In the panel (a), $W(\bfy)$ 
assumes the circles with $R=0.25$ and $|\bfy_0|=0.25 ~{\rm (solid ~curve)},
~0.75{\rm ~(dotted ~curve)},~1.25{\rm ~(dashed ~curve)}$, respectively.
These sources have the same value $|y_{\rm max}-y_{\rm min}|=0.5$. 
{}In this case, from Eq.~(\ref{condtwo}), 
the condition that the interference signature
disappears is $w\gtrsim 2\pi$.
One can confirm that the interference signature
disappears at $w\gtrsim 2\pi$ from the panel (a) of 
Figure \ref{m3_1}. 
Similarly, the panel (b) assumes the circles with the radius 
$R=0.5{\rm ~(solid ~curve)},1.0{\rm ~(dotted ~curve)}$, 
where Eq.~(\ref{condtwo}) gives 
$w\gtrsim \pi$. 
Thus, Eq.~(\ref{condtwo}) is regarded as the condition 
that the interference signature disappears due to the 
phase cancellation.

On the other hand, the panel (c) assumes the tracks of a point source
which connect the two points between $(0.01,0)~{\rm and}~(0.51,0)$ {\rm (solid curve)}, 
$(1.0,0)~{\rm and}~(1.5,0)$ {\rm (dotted ~curve)}, and $(0.5,0)~{\rm and}~(1,0)$
 {\rm (dashed curve)}. 
All these cases have $|y_{\rm max}-y_{\rm min}|=0.5$, and 
Eq.~(\ref{condtwo}) yields $w\gtrsim 2\pi$.
Similarly, the panel (d) assumes the tracks of a point source
which connect the two points, $(0.4,0)~{\rm and}~(0.4,\sqrt{1.1^2-0.4^2})$ {\rm (solid curve)},
$(0.8,0)~{\rm and}~(0.8,\sqrt{1.5^2-0.8^2})$ {\rm (dotted curve)}, and
$(1.2,0)~{\rm and}~(1.2,\sqrt{1.9^2-1.2^2})$ {\rm (dashed curve)}.
All these cases have $|y_{\rm max}-y_{\rm min}|=0.7$, and
Eq.~(\ref{condtwo}) yields $w\gtrsim 10\pi /7$.
In these cases, the interference signature slightly remains even for 
$w|y_{\rm max}-y_{\rm min}|\gtrsim\pi$, but becomes very weak there. 

In the reference \cite{MY}, it is discussed that the finite 
source size effect becomes substantial and the interference 
signature is affected under the condition $wR\gg1$.
Eq.~(\ref{condtwo}) is essentially same as the 
condition discussed in the previous paper. 

\section{Discussion}

Here let us discuss the astrophysical condition that 
the finite source size effect becomes important
in the attolensing. From the condition (\ref{condtwo}),
we have
\begin{eqnarray} 
 \hat R (1+z_L)\omega\sqrt{4GM D_L\over D_{LS}D_S} \simgt \pi, 
\label{ab}
\end{eqnarray}
where we introduced $\hat R(=|y_{\rm min}-y_{\rm max}|\theta_E D_S)$, 
which is the physical size of the source, 
and the redshift of the lens is taken into account. 
Here we imagine the black holes spread throughout 
the Universe, and consider the attolensing 
at the cosmological distance. In this case 
Eq. (\ref{ab}) is rephrased as
\begin{eqnarray} 
1.1 \times (1+z_L)\biggl({\hbar \omega \over {\rm 100MeV}}\biggr)
\biggl({M\over 10^{-19}{M_{\rm sun}}}\biggr)^{1/2}
\biggl({\hat R\over 10^{3}{\rm km}}\biggr)
\biggl({H_0^{-1}\over D_{LS}D_S/D_L}\biggr)^{1/2} \simgt \pi.
\end{eqnarray}
Note that $10^{-19} M_{\rm sun}=0.2\times 10^{15}{\rm g}$.
Thus the finite source size effect will be influential to the attolensing
by the black hole at the cosmological distances, for example, 
$\hat R$ must be less than $10^3$ km. 
However, for the attolensing by the black hole at the galactic 
scale $D_L \sim 10 {\rm~kpc}$ and $D_{LS}\sim D_S\sim H_0^{-1}$, 
the condition is relaxed by the factor 
$\sqrt{D_S/D_L}\sim 10^3$. 

When the source has the relative velocity to the lens, 
$v_{\scriptscriptstyle \bot}$, perpendicular to 
the line of sight direction, we may write
\begin{eqnarray}
  |y_{\rm min}-y_{\rm max}|\simeq{v_{\scriptscriptstyle \bot}\Delta t
\over D_S\theta_E },
\end{eqnarray}
where $\Delta t$ is the duration time of the source emission. 
{}From the condition (\ref{condtwo}),
\begin{eqnarray}
 (1+z_L)\omega\sqrt{4GM D_L\over D_{DS}D_S} v_{\scriptscriptstyle\bot}\Delta t\gtrsim\pi.
\end{eqnarray}
Then, the condition (\ref{condtwo}) is written as 
\begin{eqnarray}
  0.3\times (1+z_L) 
  \biggl({\hbar \omega \over 100{\rm MeV}}\biggr)
  \biggl({M \over 10^{-19} {M_{\rm sun}}}\biggr)^{1/2}
  \biggl({\Delta t \over 1{\rm s}}\biggr)
  \biggl({v_{\scriptscriptstyle \bot}\over 3\times 10^2{\rm km}}\biggr)
  \biggl({{H_0^{-1}}\over D_{S}D_{LS}/D_L}\biggr)^{1/2}
  \simgt \pi.
\end{eqnarray}
Thus, the relative motion of the source will be also 
influential to the attolensing by the black hole at the 
cosmological distance as well as at the galactic distance. 

\section{Summary and Conclusions}
We investigated the condition that the finite source size effect and
the relative motion of the source becomes substantial in the wave
effect of the gravitational lensing. The condition is expressed
by the formula (\ref{condtwo}), whose physical meaning is that
the difference of the phase between the two light paths from 
$y_{\rm max}$ and $y_{\rm min}$ is larger than $2\pi$.
We have shown that the finite source size effect is
important in the attolensing by the black hole at the 
cosmological distance.  Also the 
relative motion of the source to the lens 
can be influential. For the attolensing by the black hole 
at the galactic distance, the constraint is relaxed by 
the factor $\sqrt{D_S/D_L}\sim 10^{3}$. 
However, we should also note that the detection of the 
attolensing is limited in practice \cite{KPIII}, even
when the finite source size is not taken into account.

In general, the signature of the interference becomes remarkable
when $w\sim 1$, 
i.e., 
\begin{eqnarray}
  w \sim 0.3\times (1+z_L)\left({h\nu\over 100{\rm MeV}}\right)
  \left({M\over 10^{-19}M_{\rm sun}}\right)\sim 1. 
\label{condfirst}
\end{eqnarray}
In the case $w\simlt 1$, the amplification due to the lensing
becomes negligible, because the wavelength of the light becomes larger
than the size of deflector. 
Combining this and the condition (\ref{condtwo}),
we may conclude that the condition for the possible observation
of the interference signature in the gravitational lensing needs 
\begin{eqnarray}
  1\simlt w \simlt {\pi \over |y_{\rm min}-y_{\rm max}|}.
\label{finalcond}
\end{eqnarray}
Therefore, this means $|y_{\rm min}-y_{\rm max}|\simlt \pi$, 
the angular size of the source must be less than the 
Einstein radius, is always necessary for a possible observation 
of the interference signature. 
It will be useful to demonstrate the region satisfying (\ref{finalcond}) clearly.
In Figure 8, the dashed line is $w=1$, and the solid line is
$w|y_{\rm min}-y_{\rm max}| =\pi$, where we fixed 
$\hat R=10^3$ km and $D_{LS}D_S/D_L=H_0^{-1}$. 
The shaded region satisfies the condition (\ref{finalcond}).

It might also be interesting to consider whether other physical system  
satisfy the condition (\ref{finalcond}) or not. 
Figure 9 shows a region satisfying the condition with the 
parameter associated with the microlensing. 
Similar to Figure 8, the shaded region satisfies the condition. 
Here the dashed line in Figure 9 is $w=1$,
and the solid line is
$w|y_{\rm min}-y_{\rm max}| =\pi$, where we fixed $\hat R=7\times 10^5$ km (solar radius) 
and $D_{LS}D_S/D_L=30$ kpc.
The point of the intersection of these two lines is
\begin{eqnarray}
&& \left({\nu\over {\rm GHz}}\right)=8.9\times 10^2 \left({\hat R\over 7\times 10^5{\rm km}}
\right)^{-2}
  \left({30{\rm kpc}\over D_{LS}D_S/D_L}\right)^{-1}
\\
&& \left({M\over M_{\rm sun}}\right)=0.9\times 10^{-8} \left({\hat R\over 7\times 10^5{\rm km}}
\right)^{2}
  \left({30{\rm kpc}\over D_{LS}D_S/D_L}\right).
\end{eqnarray}
Thus the microlensing can be a possible system that satisfies the 
condition (\ref{finalcond}). Especially, for the microlensing 
by an Earth like planet $M\sim 10^{-5} M_{\rm sun}$, the relevant 
range of the frequency is $1$ GHz $\sim 100$ GHz. 
Then, the measurement of the microlensing event through the 
frequency might be relevant to the interference signature. 
However, it will be very difficult to detect the signal
because the stars at $1\sim 100$ GHz frequency 
band at the galactic distance is very dark in general.

\vspace{0.3cm}
\begin{acknowledgments}
This work is supported in part by Grant-in-Aid for Scientific research
of Japanese Ministry of Education, Culture, Sports, Science and
Technology (Nos.~18540277, 19035007), and a grant from Hiroshima 
University. We thank T.~Yamashita, Y.~Fukazawa, R.~Yamazaki, Y.~Kojima, 
S.~Mizuno N.~Matsunaga for useful comments and conversations 
related to the topic 
in the present paper. F.~P.~Zen and B.~E.~Gunara are grateful 
to the people at the theoretical astrophysics group and the 
elementary particle physics group of Hiroshima university 
for warmest hospitality. 
\end{acknowledgments}


\newpage

\begin{figure}
\begin{center}
\includegraphics[width=9cm,height=9cm,clip]{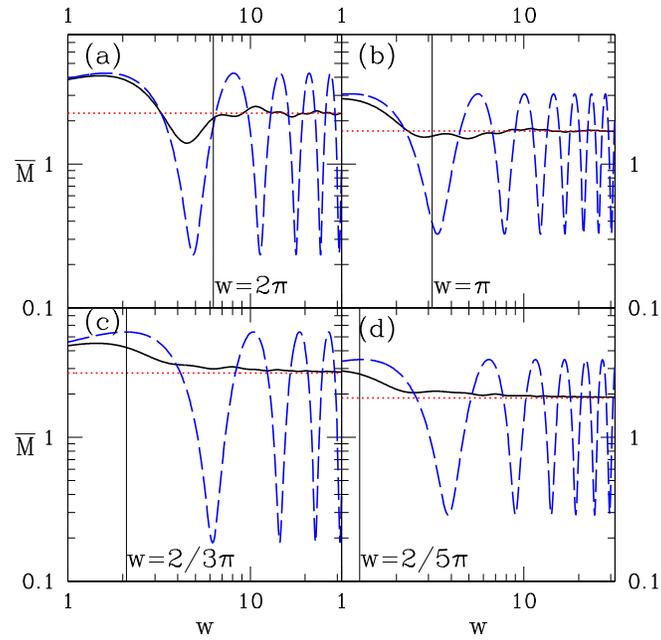}
\caption{$\bar {\cal M}(w)$ (solid curve) as function of $w$ with
$W(\bfy)$ shown in each panel in Figure 2, correspondingly.
The dashed curve is ${\cal M}(w,y_0)$, and the doted line is 
$\bar {\cal M}_{\rm geo}$.}
\label{m2_6}
\end{center}
\end{figure}
\begin{figure}
\begin{center}
\includegraphics[width=9cm,height=9cm,clip]{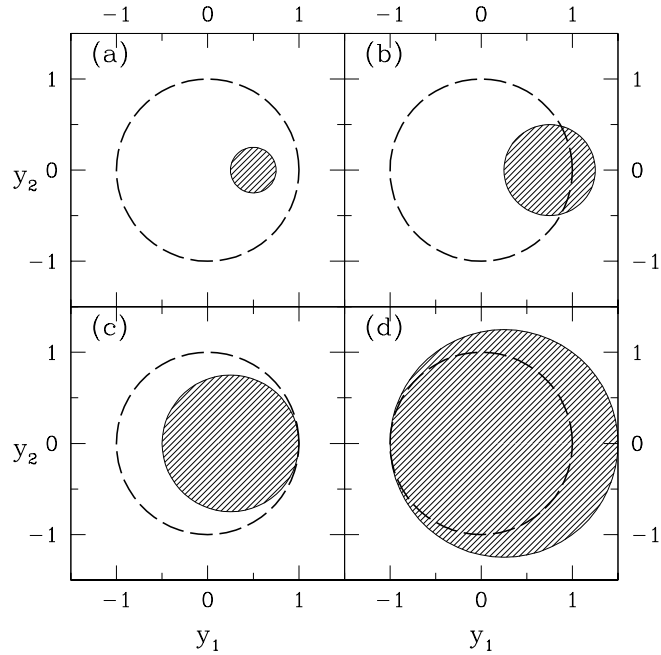}
\caption{The function $W(\bfy)$, where the region $W(\bfy)=1$ is shown by 
shaded region. The dashed circle is the Einstein radius.}
\label{m2_6w}
\end{center}
\end{figure}

\begin{figure}
\begin{center}
\includegraphics[width=9cm,height=9cm,clip]{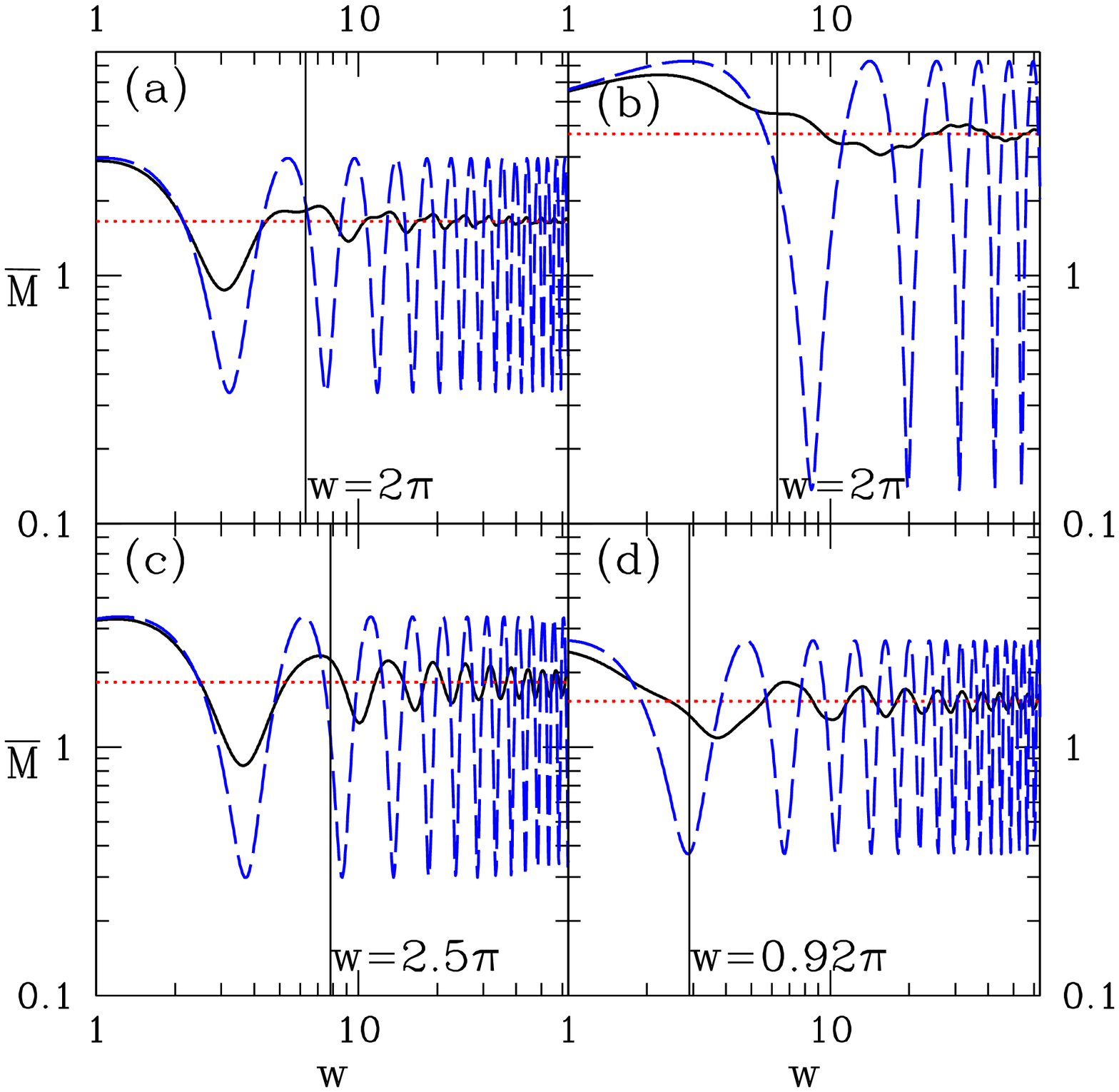}
\caption{$\bar {\cal M}(w)$ (solid curve) as function of $w$ with
$W(\bfy)$ shown in each panel in Figure 4, correspondingly.
The dashed curve is ${\cal M}(w,y_0)$, and the doted line is 
$\bar {\cal M}_{\rm geo}$.}
\label{m1_7}
\end{center}
\end{figure}

\begin{figure}
\begin{center}
\includegraphics[width=9cm,height=9cm,clip]{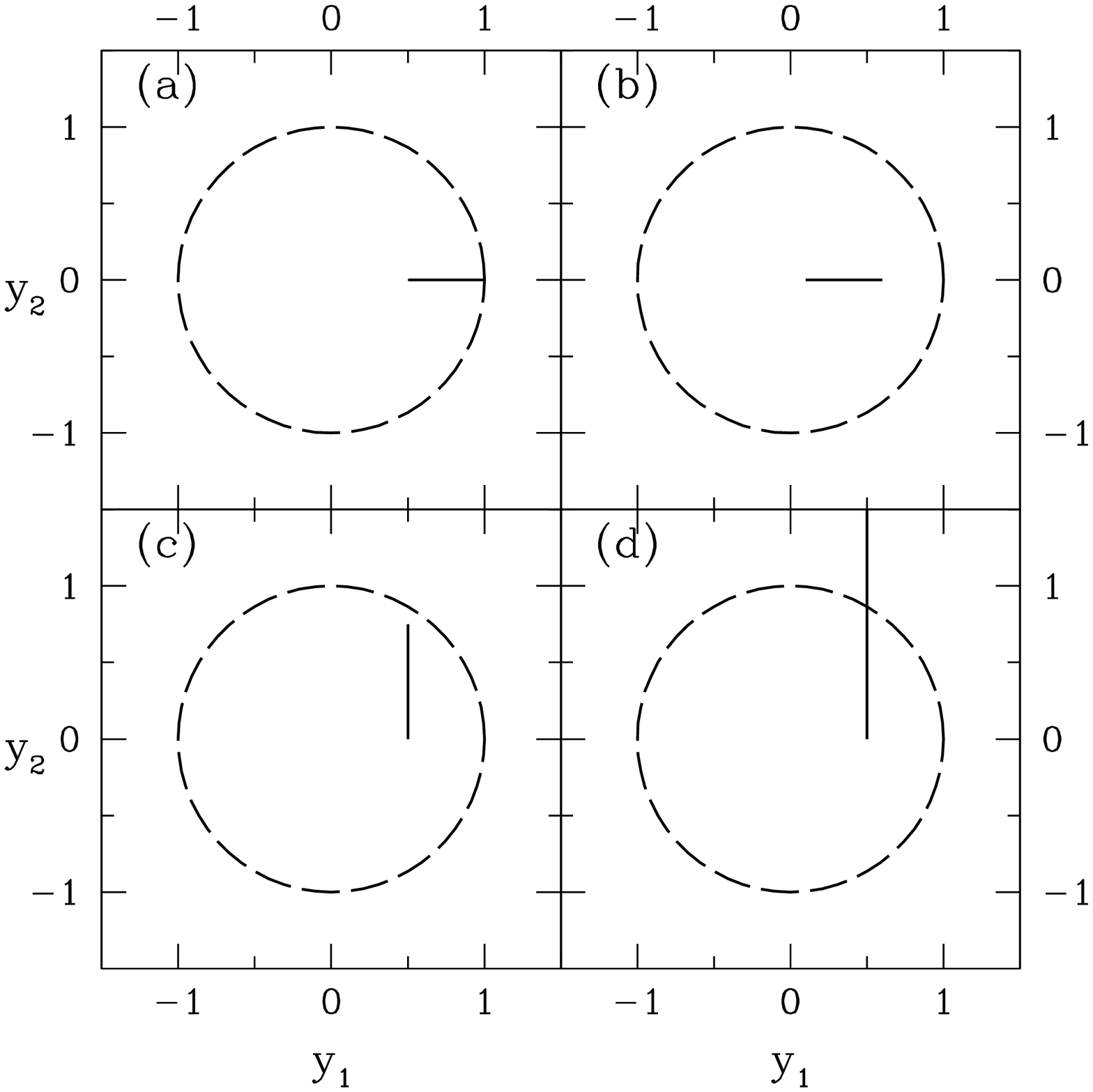}
\caption{The function $W(\bfy)$, where the region $W(\bfy)\neq1$ is shown
by straight line. The dashed circle is the Einstein radius.}
\label{m1_7w}
\end{center}
\end{figure}

\begin{figure}
\begin{center}
\includegraphics[width=9cm,height=9cm,clip]{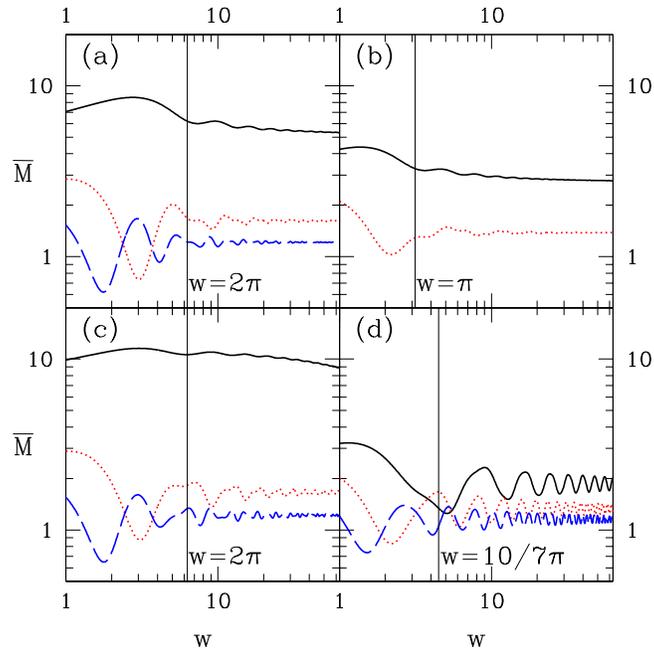}
\caption{$\bar {\cal M}(w)$ as function of $w$ with
$W(\bfy)$ shown in each panel in Figure 6, correspondingly.}
\label{m3_1}
\end{center}
\end{figure}

\begin{figure}
\begin{center}
\includegraphics[width=9cm,height=9cm,clip]{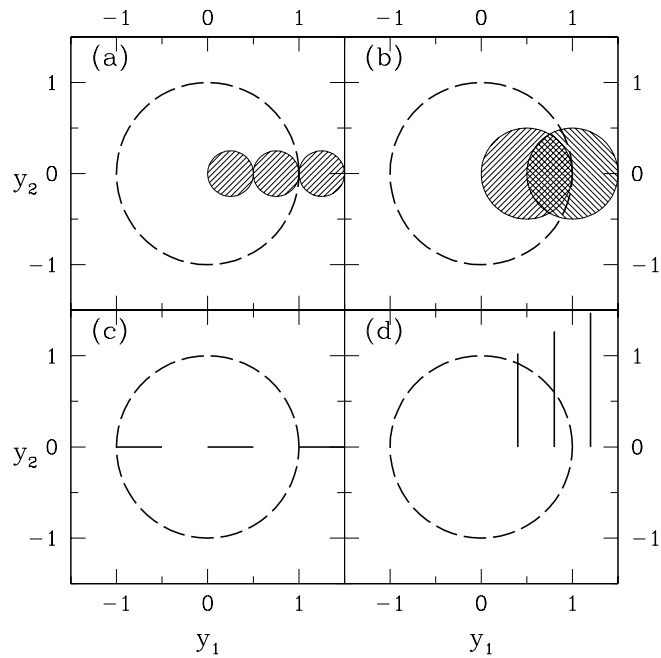}
\caption{The each panel shows $W(\bfy)$. The dashed circle is the Einstein radius.}
\label{m3_1w}
\end{center}
\end{figure}

\begin{figure}[t]
\begin{center}
\includegraphics[width=11cm,height=11cm,clip]{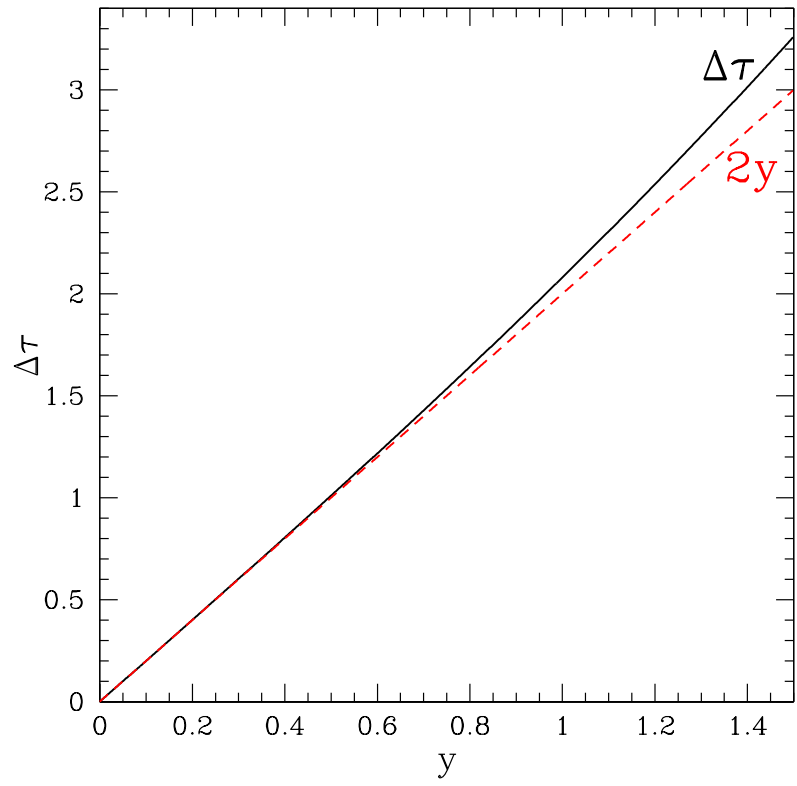}
\caption{$\tau$ (solid curve) as a function $y$. 
The dashed line is $2y$.
}
\label{dtau}
\end{center}
\end{figure}

\begin{figure}[t]
\begin{center}
\includegraphics[width=9cm,height=9cm,clip]{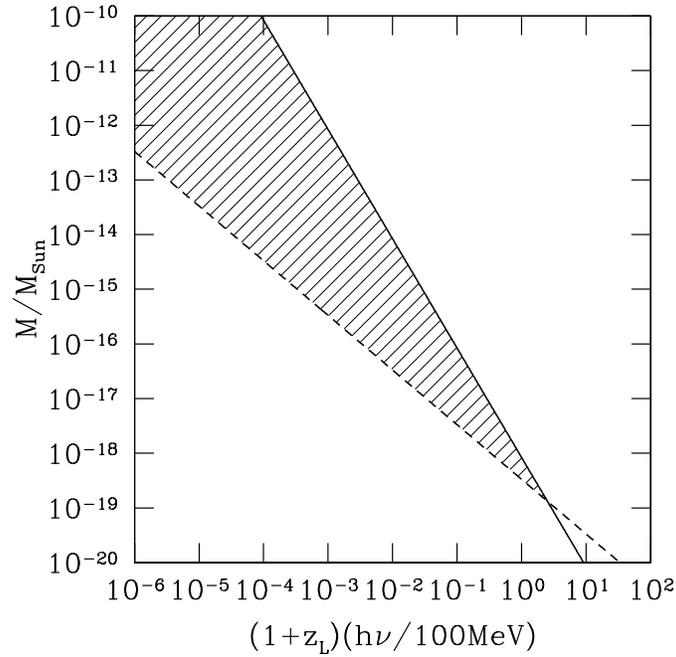}
\caption{The dashed line is $w=1$, and the solid line is
$w |y_{\rm min}-y_{\rm max}| =\pi$, where we fixed $\hat R=10^3$ km and $D_{LS}D_S/D_L=H_0^{-1}$
(cosmological distance). 
The shaded region 
satisfies the condition (\ref{finalcond}).}
\label{adda}
\end{center}
\end{figure}

\begin{figure}[t]
\begin{center}
\includegraphics[width=9cm,height=9cm,clip]{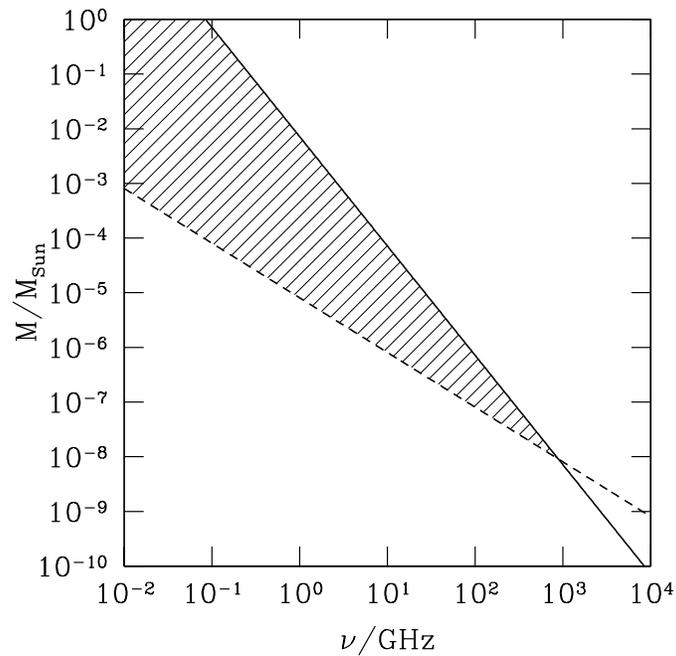}
\caption{Same as Fig.8, but we here fixed $\hat R=7\times 10^5$ km (solar radius) 
and $D_{LS}D_S/D_L=30$ kpc (galactic distance).
}
\label{addb}
\end{center}
\end{figure}

\end{document}